\documentclass[%
reprint,
showpacs,preprintnumbers,
amsmath,amssymb,
aps,
prb,
]{revtex4-1}

\usepackage{graphicx}
\usepackage{dcolumn}
\usepackage{bm}
\usepackage{colortbl}

\begin{document}


\title{Diagnostics of many-particle electronic states from non-stationary currents and residual charge}

\author{N.\,S.\,Maslova$^{1}$}
\altaffiliation{}
\author{P.\,I.\,Arseyev$^{2}$}
\author{V.\,N.\,Mantsevich$^{1}$}
\altaffiliation{} \email{vmantsev@gmail.com}

\affiliation{%
$^{1}$Moscow State University, 119991 Moscow, Russia, $^{2}$ P.N.
Lebedev Physical Institute RAS, 119991 Moscow, Russia
}%

\date{\today }
\begin{abstract}
We propose the method for identifying many particle electronic
states in the system of coupled quantum dots (impurities) with
Coulomb correlations. We demonstrate that different electronic
states can be distinguished by the complex analysis of localized
charge dynamics and non-stationary characteristics. We show that
localized charge time evolution strongly depends on the properties
of initial state and analyze different time scales in charge
kinetics for initially prepared singlet and triplet states. We
reveal the conditions for existence of charge trapping effects
governed by the selection rules for electron transitions between the
states with different occupation numbers.
\end{abstract}

 \maketitle

\section{Introduction}
The control and diagnostics of electronic states in semiconductor
nanostructures attracts a great deal of attention now a days. One of
the key problems in this area is a development of efficient methods
of detection of electronic states with different spin orientation as
spin degrees of freedom are considered to play an important role in
realizing new functions in modern nanoelectronic devices such as
spin pumps
\cite{Keller},\cite{Stoof},\cite{Covington},\cite{Mantsevich},\cite{Rezoni}
and turnstiles \cite{Pekola}, \cite{Averin}, spin interference
devices \cite{Qian}, quantum dot spin cellular automata
\cite{Bayat},\cite{Shulman},\cite{Lent} and devices for the qubit
information \cite{Hanson},\cite{Weperen}.

Double QDs are recently an attractive objects for spin-dependent
transport analysis
\cite{Golovach},\cite{Hornberger},\cite{Weymann},\cite{Mantsevich_1}.
Electronic transport through the coupled QDs was considered both in
the case of coupling to spin-polarized magnetic \cite{Hornberger}
and non-magnetic \cite{Mantsevich_1} leads. Coupled QDs can be
applied for modern nanoelectronic devices creation due to the
particular properties of charge and spin kinetics of individual
localized states
\cite{Stafford_0},\cite{Hazelzet},\cite{Cota},\cite{Mantsevich_2}.
The possibility of QDs integration in a small size quantum circuits
deals with careful analysis of relaxation processes and
non-stationary effects influence on the electron transport through
the dots system
\cite{Angus},\cite{Grove-Rasmussen},\cite{Moriyama},\cite{Landauer},\cite{Loss}.
Electronic transport in such systems is strongly governed by the
presence of Coulomb correlations and by the ratio between the QDs
coupling and interaction with the reservoir \cite{Mantsevich_3}.
Correct interpretation of quantum effects in nanoscale systems
provides an opportunity to use them as a basis for high speed
electronic and logic devices creation \cite{Tan}. Consequently, the
problem of charge kinetics in correlated low-dimensional systems due
to the coupling with reservoir is really vital. Moreover,
non-stationary characteristics provide more information about the
properties of nanoscale systems comparing to the stationary one.

In the present paper we propose the way of different many particle
electronic states characterization in the system of two interacting
quantum dots (impurity atoms) with Coulomb correlations by means of
non-stationary current analysis and investigation of the charge
trapping effects. Initial charge time evolution is analyzed in terms
of pseudo particle technique with additional constraint on possible
states
\cite{Coleman},\cite{Coleman_1},\cite{Wingreen},\cite{Barnes}.

\section{Theoretical model}

\begin{figure} [h]
\includegraphics[width=60mm]{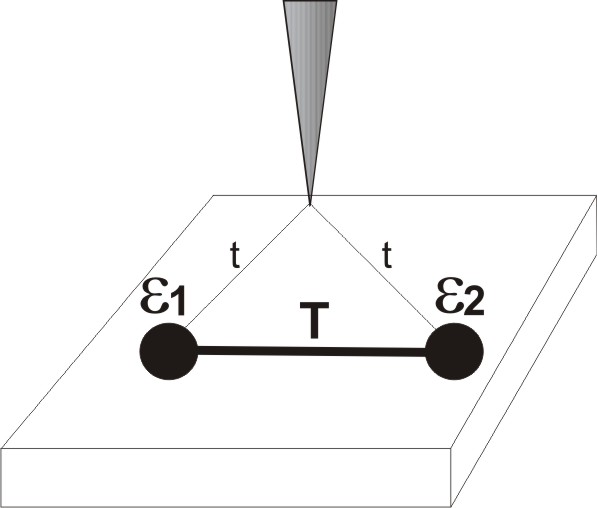}%
\caption{Sketch of two interacting quantum dots on the substrate
coupled to reservoir.} \label{figure10}
\end{figure}

We consider the problem of different many particle electronic states
resolution in the system of two interacting correlated quantum dots
(impurities) by analyzing its non-stationary characteristics.
Coupled quantum dots are weakly connected to the substrate, so there
is no charge transfer from the dots to the substrate. Charge
transfer in the system is allowed only between the dots and due to
the QDs coupling to the reservoir, switched to the system at $t=0$
(see Fig.\ref{figure10}).

The Hamiltonian of the system

\begin{eqnarray}
\hat{H}=\hat{H}_{dot}+\hat{H}_{res}+\hat{H}_{tun}
\end{eqnarray}

is written as a sum of the QDs Hamiltonian

\begin{eqnarray}
\hat{H}_{dot}=\sum_{l,\sigma}\varepsilon_{l}\hat{n}_{l\sigma}+\sum_{l,\sigma}U_{l}\hat{n}_{l\sigma}\hat{n}_{l-\sigma}+T(\hat{c}_{1\sigma}^{+}\hat{c}_{2\sigma}+\hat{c}_{2\sigma}^{+}\hat{c}_{1\sigma}),\nonumber\\
\end{eqnarray}

electronic reservoir Hamiltonian

\begin{eqnarray}
\hat{H}_{res}=\sum_{k\sigma}\varepsilon_{k}\hat{c}_{k\sigma}^{+}\hat{c}_{k\sigma}
\end{eqnarray}

and the tunneling part, which describes transitions between the dots
and reservoir

\begin{eqnarray}
\hat{H}_{tun}=
\sum_{k\sigma}t_{k1}(\hat{c}_{k\sigma}^{+}\hat{c}_{1\sigma}+\hat{c}_{1\sigma}^{+}\hat{c}_{k\sigma})+
\sum_{p\sigma}t_{k2}(\hat{c}_{k\sigma}^{+}\hat{c}_{2\sigma}+\hat{c}_{2\sigma}^{+}\hat{c}_{k\sigma}).\label{Ht}\nonumber\\
\end{eqnarray}

Here index $k$ labels continuous spectrum states in the reservoir.
Localized charge time evolution depends on the way of coupling to
reservoir as it was shown in \cite{Mantsevich_2}. We'll consider the
symmetric coupling to reservoir and assume hopping amplitudes
between the reservoir and QD with the energy $\varepsilon_l$ to be
independent on the momentum and spin, so further $t_{k1}=t_{k2}=t$.
Tunneling transfer amplitude between the dots $T$ is also considered
to be independent on the momentum and spin. Operators
$\hat{c}_{k}^{+}/\hat{c}_{k}$ are the creation/annihilation
operators for the electrons in the continuous spectrum states $k$.
$\hat{n}_{l\sigma(-\sigma)}=\hat{c}_{l\sigma(-\sigma)}^{+}\hat{c}_{l\sigma(-\sigma)}$-localized
state electron occupation numbers, where operator
$\hat{c}_{d\sigma(-\sigma)}$ destroys electron with the spin
$\sigma(-\sigma)$ on the single particle energy level
$\varepsilon_l$. $U_l$ is the on-site Coulomb repulsion for the
double occupation of the localized state. We consider low
temperature regime when Fermi level is well defined and temperature
is much lower than all the typical energy scales in the system. Let
us further consider $\hbar=1$ and $e=1$ elsewhere. As we consider
the strong coupling between the QDs, the basis of exact
eigenfunctions and eigenvalues of coupled QDs without interaction
with the reservoir should be applied. Wave functions for the single-
and multi-electronic states are well known:

Two single electron states with the wave function

\begin{eqnarray}
\Psi_{i}^{\sigma}=\mu_{i}\cdot|0\uparrow\rangle|00\rangle+\nu_{i}\cdot|00\rangle|0\uparrow\rangle.
\end{eqnarray}

and energies

\begin{eqnarray}
\varepsilon_{a(s)}=\frac{\varepsilon_1+\varepsilon_2}{2}\pm\sqrt{\frac{(\varepsilon_1+\varepsilon_2)^{2}}{4}+T^{2}}
\end{eqnarray}

exist in the system. Coefficients $\mu_{i}$ and $\nu_{i}$ are
determined by the eigenvector of matrix:

\begin{eqnarray}
\begin{pmatrix}\varepsilon_{1} && -T\\
-T && \varepsilon_{2}\end{pmatrix}. \label{m1}\end{eqnarray}

Six two electronic states exist in the system: two states with the
same spin direction $T^{+}$ $|\uparrow\uparrow\rangle$; $T^{-}$
$|\downarrow\downarrow\rangle$ and four states with the opposite
spins and wave function:

\begin{eqnarray}
\Psi_{j}^{\sigma-\sigma}&=&\alpha_{j}\cdot|\uparrow\downarrow\rangle|00\rangle+\beta_{k}\cdot|\downarrow0\rangle|0\uparrow\rangle+\nonumber\\&+&
\gamma_{j}\cdot|0\uparrow\rangle|\downarrow0\rangle+\delta_{j}\cdot|00\rangle|\uparrow\downarrow\rangle.\label{1}\nonumber\\
\end{eqnarray}

Two electron energies and coefficients $\alpha_{j}$, $\beta_{j}$,
$\gamma_{j}$ and $\delta_{j}$ are determined by the eigenvalues and
eigenvectors of matrix:

\begin{eqnarray}
\begin{pmatrix}2\varepsilon_{1}+U_{1} && -T && -T && 0 \\
-T && \varepsilon_{1}+\varepsilon_{2} && 0 && -T\\
-T && 0 && \varepsilon_{1}+\varepsilon_{2} && 0\\
0 && -T && -T && 2\varepsilon_{2}+U_{2}\end{pmatrix}.
\label{m2}\end{eqnarray}

These are low energy singlet $S^{0}$ and triplet $T^{0}$ states and
excited singlet and triplet states $S^{0*}$ and $T^{0*}$. Low energy
triplet state $T^{0}$ with energy $\varepsilon_1+\varepsilon_2$
exists for any values of QDs energy levels $\varepsilon_l$ and
Coulomb interaction $U_l$. Corresponding coefficients in
Eq.(\ref{1}) are $\alpha=\delta=0$ and
$\beta=-\gamma=\frac{1}{\sqrt{2}}$.

Two three electron states with the wave function

\begin{eqnarray}
\Psi_{m}^{\sigma\sigma-\sigma}&=&p_{m}|\uparrow\downarrow\rangle|\uparrow\rangle+q_{m}|\uparrow\rangle|\uparrow\downarrow\rangle\nonumber\\
m&=&\pm1
\end{eqnarray}

exist in the system. Coefficients $p_{m}$ and $q_{m}$ and energies
are determined by the eigenvectors and eigenvalues of matrix:

\begin{eqnarray}
\begin{pmatrix}2\varepsilon_{1}+\varepsilon_{2}+U_{1} && -T\\
-T && 2\varepsilon_{2}+\varepsilon_{1}+U_{2}\end{pmatrix}
\label{m3}\end{eqnarray}

Finally, single four-electronic state exists in the system with the
wave function

\begin{eqnarray}
\Psi_{n}=|\uparrow\downarrow\rangle|\uparrow\downarrow\rangle.
\end{eqnarray}

Coupling to reservoir leads to the changing of electrons number in
the dots due to the tunneling processes. Kinetic properties are
governed by the selection rules, which are determined by the matrix
elements between the states with different number of electrons.
Transitions between the states with different number of electrons
can be analyzed in terms of pseudo-particle operators with
constraint on the physical states (the number of pseudo-particles).
Consequently, the electron operator $c_{\sigma l}^{+}$  $(l=1,2)$
can be written in terms of pseudo-particle operators:

\begin{eqnarray}
c_{\sigma l}^{+}&=&\sum_{i}X_{i}^{\sigma l}f_{\sigma
i}^{+}b+\sum_{j,i}Y_{ji}^{\sigma-\sigma
l}d_{j}^{+\sigma-\sigma}f_{i-\sigma}+\\&+&\sum_{i}Y_{i}^{\sigma\sigma
l}d^{+\sigma\sigma}f_{i\sigma}+\sum_{m,j}Z_{mj}^{\sigma\sigma-\sigma
l}\psi_{m-\sigma}^{+}d_{j}^{\sigma-\sigma}+\nonumber\\&+&\sum_{m}Z_{m}^{\sigma-\sigma-\sigma
l}\psi_{m\sigma}^{+}d^{-\sigma-\sigma}+\sum_{m}W_{m}^{\sigma-\sigma-\sigma
l}\varphi^{+}\psi_{m\sigma}\nonumber\
\end{eqnarray}

where $f_{\sigma}^{+}(f_{\sigma})$ and
$\psi_{\sigma}^{+}(\psi_{\sigma})$- are pseudo-fermion creation
(annihilation) operators for the electronic states with one and
three electrons correspondingly. $b^{+}(b)$,
$d_{\sigma}^{+}(d_{\sigma})$ and $\varphi^{+}(\varphi)$- are slave
boson operators, which correspond to the states without any
electrons, with two electrons or four electrons. Operators
$\psi_{m-\sigma}^{+}$- describe system configuration with two spin
up electrons $\sigma$ and one spin down electron $-\sigma$ in the
symmetric and asymmetric states.

Matrix elements $X_{i}^{\sigma l}$, $Y_{ji}^{\sigma-\sigma l}$,
$Y_{ji}^{\sigma\sigma l}$, $Z_{mj}^{\sigma\sigma-\sigma l}$,
$Z_{mj}^{\sigma-\sigma-\sigma l}$ and $W_{m}^{\sigma-\sigma-\sigma
l}$ can be defined as:

\begin{eqnarray}
X_{i}^{\sigma l}&=&\langle\Psi_{i}^{\sigma}|c_{\sigma l}^{+}|0\rangle\nonumber\\
Y_{ji}^{\sigma-\sigma l}&=&\langle\Psi_{j}^{\sigma-\sigma}|c_{\sigma l}^{+}|\Psi_{i}^{-\sigma}\rangle\nonumber\\
Y_{ji}^{\sigma\sigma l}&=&\langle\Psi_{j}^{\sigma\sigma}|c_{\sigma l}^{+}|\Psi_{i}^{\sigma}\rangle\nonumber\\
Z_{mj}^{\sigma\sigma-\sigma l}&=&\langle\Psi_{m}^{\sigma\sigma-\sigma}|c_{\sigma l}^{+}|\Psi_{j}^{\sigma-\sigma}\rangle\nonumber\\
Z_{m}^{\sigma-\sigma-\sigma l}&=&\langle\Psi_{m}^{\sigma-\sigma-\sigma}|c_{\sigma l}^{+}|\Psi^{-\sigma-\sigma}\rangle\nonumber\\
W_{m}^{\sigma-\sigma-\sigma
l}&=&\langle\Psi_{n}^{\sigma\sigma-\sigma-\sigma}|c_{\sigma
l}^{+}|\Psi_{m}^{\sigma-\sigma-\sigma}\rangle\
\end{eqnarray}

Finally one can easily express matrix elements through the matrixes
(\ref{m1}), (\ref{m2}), (\ref{m3}) eigenvectors elements:

\begin{eqnarray}
X_{i}^{\sigma 1}=\mu_{i};
X_{i}^{\sigma 2}=\nu_{i}\nonumber\\
Y_{ji}^{\sigma-\sigma 1}=\alpha_j\mu_i+\beta_j\nu_i\nonumber\\
Y_{ji}^{\sigma-\sigma 2}=\delta_j\nu_i+\gamma_j\mu_i\nonumber\\
Y_{ji}^{\sigma\sigma 1}=\nu_i;
Y_{ji}^{\sigma\sigma 2}=\mu_i\nonumber\\
Z_{mj}^{\sigma\sigma-\sigma 1}=p_m\gamma_j+q_m\delta_j\nonumber\\
Z_{mj}^{\sigma\sigma-\sigma 2}=p_m\alpha_j+q_m\beta_j\nonumber\\
Z_{mj}^{\sigma-\sigma-\sigma 1}=p_m;
Z_{mj}^{\sigma-\sigma-\sigma 1}=q_m\nonumber\\
W_{m}^{\sigma-\sigma-\sigma 1}=q_m; W_{m}^{\sigma-\sigma-\sigma
2}=p_m
\end{eqnarray}

For identical QDs and arbitrary values of Coulomb correlations
charge trapping can occur due to the presence of one particle
$"$dark state$"$ with the energy $\varepsilon+T$. Matrix element,
which corresponds to transitions between this state and empty states
is equal to zero. This transition is forbidden by the symmetry of
the tunneling Hamiltonian (\ref{Ht}). If there is the way of allowed
transitions from initial state to the $"$dark state$"$ during
relaxation processes, the residual charge is trapped in this
one-particle state.

Conditions, which determine allowed and restricted transitions
between the states with different number of electrons can be easily
found:

I. Transitions between the state with zero electrons and single
electron state:

\begin{eqnarray}
X_{i}^{\sigma}=\sum_{l=1,2}X_{i}^{\sigma
l}=\sum_{l=1,2}\langle\psi_{i}^{\sigma}|c_{\sigma
l}^{+}|0\rangle=\mu_{i}+\nu_{i}=\nonumber\\=\{\begin{array}{ccccc}
0 & restricted\\
\neq0 & allowed\\
\end{array}
\end{eqnarray}

II. Transitions between single electron state and two electron state
with:

\begin{eqnarray}
Y_{ji}^{\sigma-\sigma}=\sum_{l=1,2}Y_{ji}^{\sigma-\sigma
l}=\sum_{l=1,2}\langle\psi_{j}^{\sigma-\sigma}|c_{\sigma
l}^{+}|\psi_{i}^{-\sigma}\rangle=\nonumber\\=\alpha_j\mu_i+\beta_j\nu_i+\delta_j\nu_i+\gamma_j\mu_i=\nonumber\\=\{\begin{array}{ccccc}
0 & restricted\\
\neq0 & allowed\\
\end{array}
\end{eqnarray}

and

\begin{eqnarray}
Y_{ji}^{\sigma\sigma}=\sum_{l=1,2}Y_{ji}^{\sigma\sigma
l}=\sum_{l=1,2}\langle\psi_{j}^{\sigma\sigma}|c_{\sigma
l}^{+}|\psi_{i}^{\sigma}\rangle=\mu_{i}+\nu_{i}=\nonumber\\=\{\begin{array}{ccccc}
0 & restricted\\
\neq0 & allowed\\
\end{array}
\end{eqnarray}

III. Transitions between two electron state and three electron
state:

\begin{eqnarray}
Z_{mj}^{\sigma\sigma-\sigma}=\sum_{l=1,2}Z_{mj}^{\sigma\sigma-\sigma
l}=\sum_{l=1,2}\langle\psi_{m}^{\sigma\sigma-\sigma}|c_{\sigma
l}^{+}|\psi_{j}^{\sigma-\sigma}\rangle=\nonumber\\p_m\gamma_j+q_m\delta_j+p_m\alpha_j+q_m\beta_j=\nonumber\\=\{\begin{array}{ccccc}
0 & restricted\\
\neq0 & allowed\\
\end{array}
\end{eqnarray}

and

\begin{eqnarray}
Z_{mj}^{\sigma-\sigma-\sigma}=\sum_{l=1,2}Z_{mj}^{\sigma-\sigma-\sigma
l}=\sum_{l=1,2}\langle\psi_{m}^{\sigma-\sigma-\sigma}|c_{\sigma
l}^{+}|\psi^{-\sigma-\sigma}\rangle=\nonumber\\=p_m+q_m=\{\begin{array}{ccccc}
0 & restricted\\
\neq0 & allowed\\
\end{array}
\nonumber\\
\end{eqnarray}

IV. Transitions between three electron state and four electron
state:

\begin{eqnarray}
W_{m}^{\sigma-\sigma-\sigma}=\sum_{l=1,2}W_{m}^{\sigma-\sigma-\sigma
l}=\sum_{l=1,2}\langle\psi_{l}^{\sigma\sigma-\sigma-\sigma}|c_{\sigma
l}^{+}|\psi_{m}^{\sigma-\sigma-\sigma}\rangle\
=\nonumber\\=p_m+q_m=\{\begin{array}{ccccc}
0 & restricted\\
\neq0 & allowed\\
\end{array}
\nonumber\\
\end{eqnarray}

For slightly different QDs ($\varepsilon_1\neq\varepsilon_2$) these
matrix elements determine different time scales of the system
dynamics. For rather large values of Coulomb interaction $U_{l}$ and
low temperatures only single electron and low energy two-electron
states can be considered, as all other states are separated by the
Coulomb gap. So, the following non-stationary system of equations
can be obtained for the pseudo particle filling numbers $N_{i}$,
$N_{dj}^{\sigma-\sigma}$, $N_{dj}^{\sigma\sigma}$ and $N_{b}$:

\begin{eqnarray}
\frac{\partial N_{dj}^{\sigma-\sigma}}{\partial
t}&=&-2\gamma\sum_{i\sigma
}|Y_{ji}^{\sigma-\sigma}|^{2}\cdot N_{dj}^{\sigma-\sigma}\nonumber\\
\frac{\partial N_{i\sigma}}{\partial t}&=&2\gamma\sum_{j\sigma
}|Y_{ji}^{\sigma-\sigma}|^{2}N_{dj}^{\sigma-\sigma}-|X_{i}^{\sigma}|^{2}N_{i}^{\sigma}+\sum_{j
}|Y_{ji}^{\sigma\sigma}|^{2}\cdot N_{dj}^{\sigma\sigma}\nonumber\\
\frac{\partial N_{b}}{\partial t}&=&2\gamma\sum_{i\sigma
}|X_{i}^{\sigma}|^{2}\cdot N_{i}^{\sigma}\nonumber\\
\frac{\partial N_{dj}^{\sigma\sigma}}{\partial
t}&=&-2\gamma\sum_{i\sigma }|Y_{ji}^{\sigma\sigma}|^{2}\cdot
N_{dj}^{\sigma\sigma} \label{sys1}
\end{eqnarray}

where

\begin{eqnarray}
|X_{i}^{\sigma}|^{2}&=&|\nu_i+\mu_i|^{2},\nonumber\\
|Y_{ji}^{\sigma-\sigma}|^{2}&=&|\alpha_j\mu_i+\beta_j\nu_i+\gamma_j\mu_i+\delta_j\nu_i|^{2},\nonumber\\
|Y_{ji}^{\sigma\sigma}|^{2}&=&|\nu_i+\mu_i|^{2}
\end{eqnarray}

and $\gamma=\nu_{0}t^{2}$, $\nu_{0}$ - is the unperturbed density of
states in the reservoir. Depending on the tunneling barrier width
and height typical tunneling coupling strength $\gamma$ can vary
from $10\mu eV$ \cite{Amaha} to and $1\div5$ meV\cite{Fransson}.

System of Eqs.(\ref{sys1}) for the single and two-electron states
can be solved both numerically and analytically with initial
conditions $N_{dj}^{\sigma\sigma^{'}}(0)=1$, $N_{a\sigma}(0)=0$,
$N_{s\sigma}(0)=0$ and $N_{b\sigma}(0)=0$. Analytical expressions,
which determine charge relaxation have the following form:

\begin{eqnarray}
N_{dj}^{\sigma-\sigma}(t)&=&e^{-2\lambda t},\nonumber\\
N_{a\sigma}(t)&=&\frac{\lambda_{ja}}{2\lambda-\lambda_a}\cdot(e^{-\lambda_at}-e^{-2\lambda
t}),\nonumber\\
N_{s\sigma}(t)&=&\frac{\lambda_{js}}{2\lambda-\lambda_s}\cdot(e^{-\lambda_st}-e^{-2\lambda
t}),\nonumber\\
N_{b}(t)&=&1-N_{dj}^{\sigma-\sigma}(t)-\sum_\sigma
N_{a\sigma}(t)-\sum_\sigma n_{s\sigma}(t)
\end{eqnarray}

where

\begin{eqnarray}
\lambda&=&2\gamma\cdot\sum_{i}|\alpha_j\mu_i+\beta_j\nu_i+\delta_j\nu_i+\gamma_j\mu_i|^{2},\nonumber\\
\lambda_{a}&=&2\gamma\cdot|\mu_a+\nu_a|^{2},\nonumber\\
\lambda_{s}&=&2\gamma\cdot|\mu_s+\nu_s|^{2},\nonumber\\
\lambda_{ja}&=&2\gamma\cdot|\alpha_j\mu_a+\beta_j\nu_a+\delta_j\nu_a+\gamma_j\mu_a|^{2},\nonumber\\
\lambda_{js}&=&2\gamma\cdot|\alpha_j\mu_s+\beta_j\nu_s+\delta_j\nu_s+\gamma_j\mu_s|^{2}.
\end{eqnarray}

Index $a$ corresponds to the state with energy:

\begin{eqnarray}
\varepsilon_{a}=\frac{\varepsilon_1+\varepsilon_2}{2}+\sqrt{\frac{(\varepsilon_1+\varepsilon_2)^{2}}{4}+T^{2}}
\end{eqnarray}

and index $s$ - to the state with energy:

\begin{eqnarray}
\varepsilon_{s}=\frac{\varepsilon_1+\varepsilon_2}{2}-\sqrt{\frac{(\varepsilon_1+\varepsilon_2)^{2}}{4}+T^{2}}
\end{eqnarray}

Electron occupation numbers $N_{el}$ can be obtained from the pseudo
particle occupation numbers considering spin degrees of freedom by
the following expression:

\begin{eqnarray}
N_{el}(t)= 2\cdot N_{dj}^{\sigma-\sigma}(t)+\sum_\sigma
N_{a\sigma}(t)+ \sum_\sigma N_{s\sigma}(t)\nonumber\\
\end{eqnarray}

We'll consider charge time evolution from the singlet and triplet
initial states. For singlet initial state coefficients $\alpha$,
$\beta$, $\gamma$ and $\delta$ are determined as an eigenvector of
matrix (\ref{m2}) corresponding to its minimal eigenvalue.

For the triplet initial state coefficients $\alpha=\delta=0$ and
$\beta=-\gamma=\frac{1}{\sqrt{2}}$.

\begin{figure} [h]
\includegraphics[width=80mm]{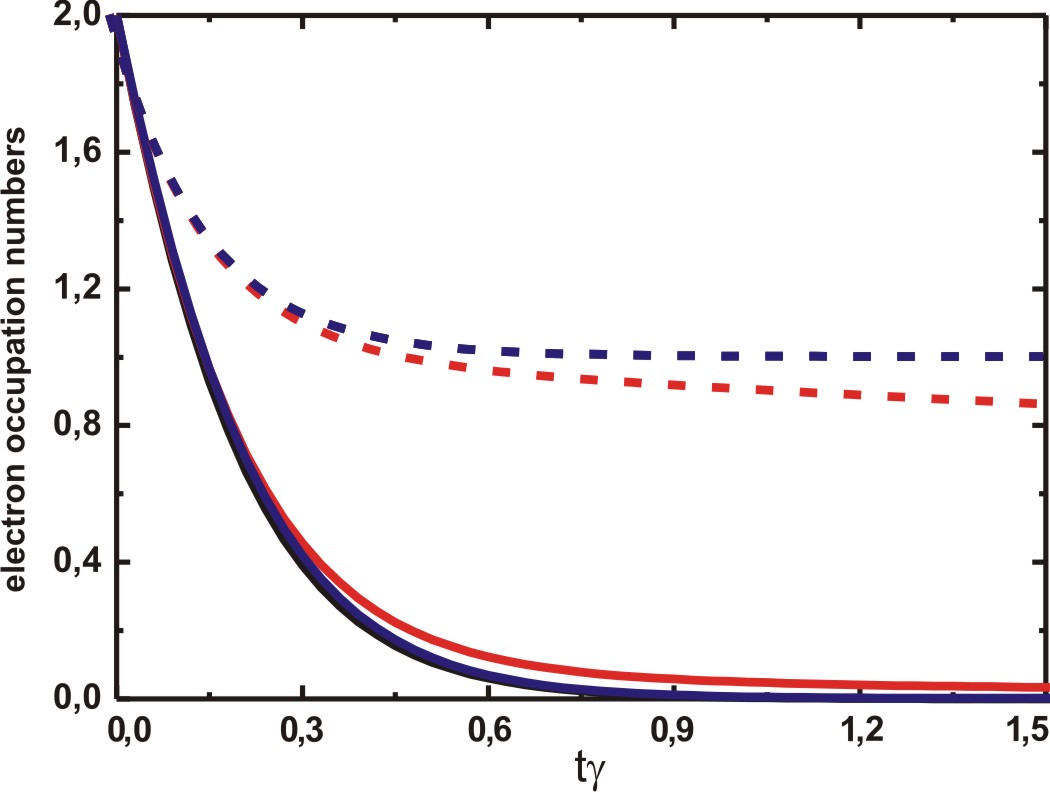}%
\caption{(Color online) Electron occupation numbers time evolution
from initial singlet state - solid lines and triplet state - dashed
line. Black solid and dashed lines:
$\varepsilon_1/\gamma=\varepsilon_2/\gamma=7$ and
$U_1/\gamma=U_2/\gamma=20$; red solid and dashed lines:
$\varepsilon_1/\gamma=7.4$, $\varepsilon_2/\gamma=7$ and
$U_1/\gamma=U_2/\gamma=20$; blue solid and dashed lines:
$\varepsilon_1/\gamma=\varepsilon_2/\gamma=7$ and $U_1/\gamma=21$,
$U_2/\gamma=20$. Parameters $T/\gamma=2$ and $\gamma=1$ are the same
for all the figures.} \label{figure20}
\end{figure}

Non-stationary behavior of the system occupation numbers depends on
the initial conditions. Fig.\ref{figure20} demonstrates charge
relaxation from initially occupied singlet (see solid lines in
Fig.\ref{figure20}) and triplet (see dashed lines in
Fig.\ref{figure20}) states for different system parameters.

When relaxation starts from the singlet state charge trapping
effects are not present in the system even for identical QDs
($\varepsilon_1=\varepsilon_2$ and $U_1=U_2$). Charge trapping is
present for identical QDS ($\varepsilon_1=\varepsilon_2$ and
$U_1=U_2$) when relaxation starts from the triplet state (see black
dashed line in the Fig.\ref{figure20}). This is the direct
manifestation of selections rules influence on the electrons
transitions between QDs states and reservoir, which are determined
by the matrix elements.

For different QDs, when conditions $\varepsilon_1\neq\varepsilon_2$
and $\Delta\varepsilon/T<<1$ are fulfilled, two different timescales
for charge relaxation from the states with the energies
$\varepsilon_{a}$ and $\varepsilon_{s}$ exist in the system. The
presence of initial energy levels detuning $\Delta\varepsilon$ leads
to the appearance of two time scales $\gamma_s$ and $\gamma_a$,
related by the ration:

\begin{eqnarray}
\gamma_a=\frac{\Delta\varepsilon^{2}}{T^{2}}\cdot\gamma_s.
\end{eqnarray}

All other relaxation rates, present in the system can be expressed
through $\gamma_s$ and $\gamma_a$:

\begin{eqnarray}
\gamma_{T^{0}s}=\frac{1}{2}\cdot\gamma_a,\nonumber\\
\gamma_{T^{0}a}=\frac{1}{2}\cdot\gamma_s,\nonumber\\
\gamma_{S^{0}s}=|\alpha+\beta|^{2}\cdot\gamma_s,\nonumber\\
\gamma_{S^{0}a}=|\alpha+\beta|^{2}\cdot\gamma_a.
\end{eqnarray}

Charge relaxation from the triplet state to the single electron
states does not depend on the value of Coulomb interaction $U_d$.
Moreover, relaxation rates from the single electron states also do
not depend on the value of Coulomb correlations (see
Fig.\ref{figure20}). For identical QDs time evolution of initial
triplet state leads to charge trapping. For slightly different QDs
the second relaxation time scale $\gamma_a$ appears in the system.
It reveals in the slow charge relaxation instead of charge trapping
(see red dashed line in the Fig.\ref{figure20}) obtained in the
absence of energy levels detuning. For initially singlet state the
presence of Coulomb interaction slightly changes relaxation dynamics
and do not influence the relaxation processes if initial state is a
triplet one.

\begin{figure}
\includegraphics[width=80mm]{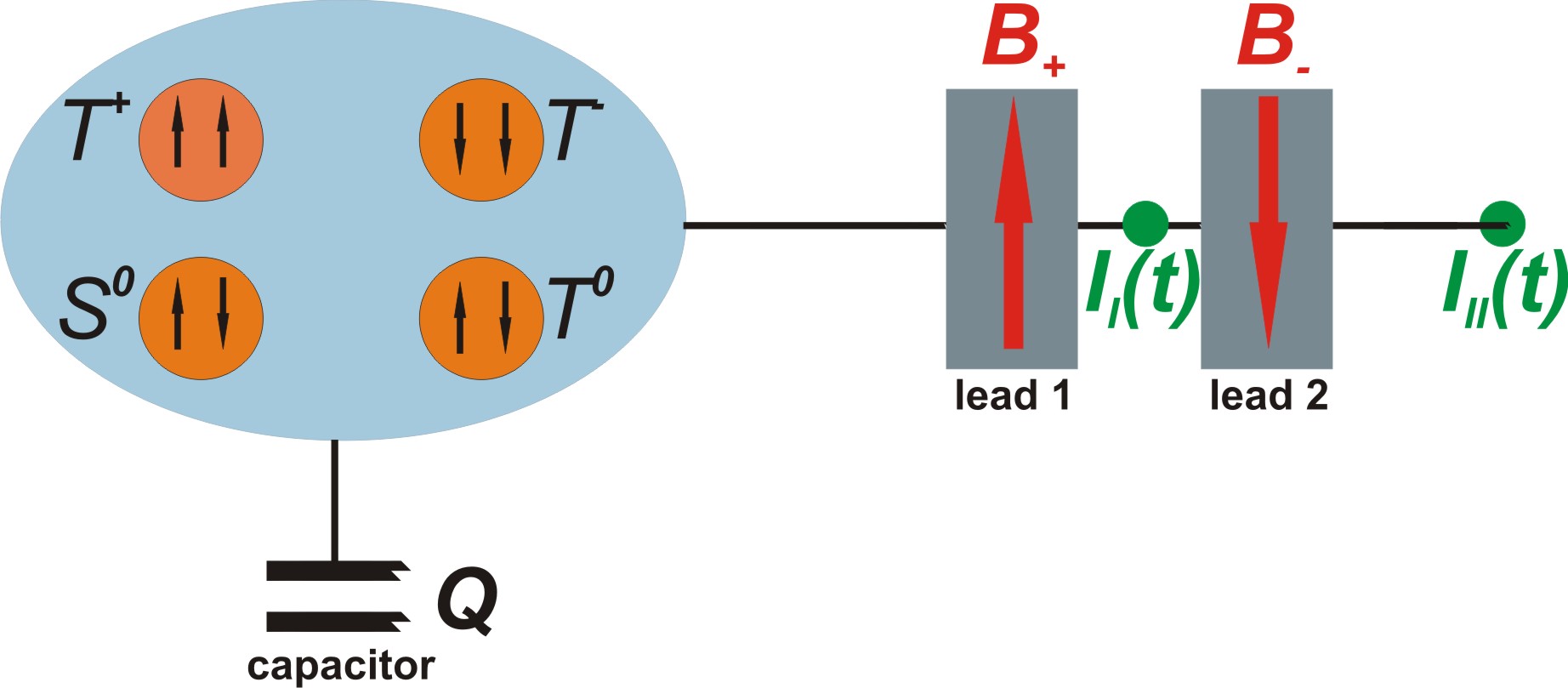}%
\caption{(Color online) Sketch of the measurement scheme, applicable
for distinguishing two-electronic states with different spin
orientation.} \label{figure5}
\end{figure}

Obtained significant difference in non-stationary behavior of
localized charge for singlet and triplet states gives us possibility
to propose experimental scheme, which allows to distinguish
different two-electronic states (see Fig.\ref{figure5}).

Let us consider the situation, when unknown initial two-electronic
state is prepared. There exists four different configurations
$S^{0}$, $T^{0}$, $T^{+}$ or $T^{-}$. To define the particular
initial state one has to analyze non-stationary system
characteristics and to control the value of residual charge $Q$. To
measure non-stationary current (see $I_{I}$ and $I_{II}$ in the
Fig.\ref{figure5}) one has to use the system of two consecutive
spin-polarized leads (see $lead_{1}$ and $lead_{2}$ in the
Fig.\ref{figure5}) with opposite directions of applied external
magnetic field (see $B_{+}$ and $B_{-}$ in the Fig.\ref{figure5}).

\begin{figure} [h]
\includegraphics[width=40mm]{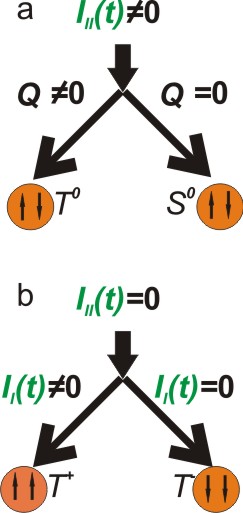}%
\caption{(Color online) Sketch of the logic diagram, which enables
to resolve initial many particle electronic states with different
spin orientation.} \label{figure6}
\end{figure}

Proposed experimental scheme allows to distinguish different
two-electronic states in the system of two interacting quantum dots
with Coulomb correlations by means of the logic diagram shown in
Fig.\ref{figure6}.

States $S^{0}$ and $T^{0}$ can be resolved by means of residual
charge $Q$ control just after non-stationary current pulse
$I_{II}\neq0$ has been registered after both leads. If measured
charge is equal to zero $Q=0$, then initial state was $S^{0}$. In
the case of non-zero charge $Q\neq0$ initial state was $T^{0}$ (see
logic diagram in the Fig.\ref{figure6}a). Initial states $T^{+}$ and
$T^{-}$ can be resolved only by the measurements of non-stationary
currents. If non-stationary current after both leads is absent
$I_{II}=0$ but non-stationary current between the leads $I_{I}$ has
non-zero value, then initial state was $T^{+}$. Contrary, the
absence of current registered between the leads $I_{I}=0$ means that
initial state was $T^{-}$ (see logic diagram in the
Fig.\ref{figure6}b).

\section{Conclusion}

We have proposed the method of how to distinguish different
two-particle electronic states in the system of two interacting
quantum dots (impurity atoms) with Coulomb correlations. This method
is based both on the analysis of the system non-stationary current
characteristics and on the control under the residual charge value.
The resolution of two-particle electronic states becomes possible
due to appearance of significantly different time scales in dynamics
of initially singlet and triplet states. We also demonstrated the
possibility of charge trapping in the proposed system due to the
selection rules presence, which govern electron transitions between
different electronic states. The method discussed in the paper can
be used for the control under many-particle electronic states in
modern nanoelectronic devices such as electronic pumps, turnstiles,
logic and quantum information devices and new types of electronic
devices based on non-equilibrium non-stationary currents.

This work was supported by RSF grant $16-12-00072$.

 \pagebreak

\end{document}